\makeatletter
\def\acmConference@name{}
\def\acmConference@shortname{}
\makeatother
\documentclass[12pt,acmlarge,screen,natbib=false,nonacm]{acmart}

\usepackage{ragged2e} 
\usepackage{etoolbox} 

\makeatletter

\renewcommand\author[2][]{
  \gdef\@author{#2}
  \ifstrequal{#1}{}
    {\gdef\@shortauthors{#2}}
    {\gdef\@shortauthors{#1}}
}

\def\maketitle{
    \vspace*{1\baselineskip}

    {\sffamily\bfseries\LARGE\noindent\color{ACMDarkBlue}\@title\par}

    \bgroup
    \sffamily\itshape
    \vspace{1\baselineskip}
    
    \noindent Technical Report RT-MAC-2021-01\\
    \noindent Department of Computer Science -- IME\\
    \noindent University of S\~ao Paulo\par
    \egroup

    \vspace{\baselineskip}

    \bgroup
    \sffamily\large
    \RaggedRightRightskip 5em plus 3em minus 3em
    \RaggedRight
    \@author\par
    \egroup

    \vspace{\baselineskip}
}

\def\customAppendixFormat{
    \clearpage

    \addtocontents{toc}{\par\vspace{1.5\baselineskip}}
    \addtocontents{toc}{\hspace{54pt}{\sffamily\bfseries Appendices\par}}
    \addtocontents{toc}{\par\vspace{.2\baselineskip}}

    \titleformat{\section}[hang]
      {\sffamily\Large\bfseries\color{ACMDarkBlue}}
      {\appendixname\space\thesection\hspace{.5em}--}
      {0.5em}
      {##1\filright}[\thickerrule]

    \titlecontents{section}[64pt]
      {\addvspace{.8\baselineskip}\sffamily\bfseries\protect\color{ACMDarkBlue}}
      {\thecontentslabel\space--\space}
      {}
      {\titlerule*{.}\thecontentspage}
      [\addvspace{.2\baselineskip}]
}

\appto\appendix{\customAppendixFormat}

\makeatother

\renewenvironment{abstract}
  {
    \begin{list}
      {}
      {
        \rightmargin 4em
        \leftmargin 0pt
        \itemindent 0pt
      }
      \small\item[]\textbf{Abstract:}\enspace
  }
  {\end{list}}

\usepackage[explicit]{titlesec}

\def\thickerrule{\titlerule[1.2pt]}

\titleformat{\section}[hang]
  {\sffamily\Large\bfseries\color{ACMDarkBlue}}
  {\thesection}
  {0.9em}
  {#1}[\thickerrule]

\titlespacing{\section}{0pt}{20pt}{33pt}

\titleformat{\subsection}[hang]
  {\sffamily\bfseries\color{ACMDarkBlue}}
  {\thesubsection}
  {0.7em}
  {#1}

\titlespacing{\subsection}{0pt}{20pt}{15pt}

\titleformat{\subsubsection}[hang]
  {\itshape\color{ACMDarkBlue}}
  {}
  {0pt}
  {\underline{\strut#1}}

\titlespacing{\subsubsection}{0pt}{25pt}{20pt}

\usepackage{titletoc}

\titlecontents{section}[64pt]
  {\addvspace{.8\baselineskip}\sffamily\bfseries\color{ACMDarkBlue}}
  {}
  {}
  {\titlerule*{.}\thecontentspage}
  [\addvspace{.2\baselineskip}]

\titlecontents{subsection}[64pt]
  {\sffamily}
  {}
  {}
  {\titlerule*{.}\thecontentspage}

\titlecontents*{subsubsection}[74pt]
  {\addvspace{.2\baselineskip}\sffamily\itshape}
  {}
  {}
  {\hspace*{8pt}\upshape\thecontentspage}[\enspace--\enspace]

\usepackage{enumitem}
\newlist{exercises}{description}{3}
\setlist[exercises]{topsep=1.5\baselineskip,itemsep=1.5\baselineskip,left=0pt,parsep=.5\baselineskip}
\setlist[enumerate]{label=\arabic*.}

\usepackage{graphicx}
\graphicspath{{images/}}

\usepackage{pdfpages}

\RequirePackage[pages=some]{background}
\backgroundsetup{scale=1,color=black,opacity=1.0,angle=0, contents={\includegraphics[width=\paperwidth,height=\paperheight]{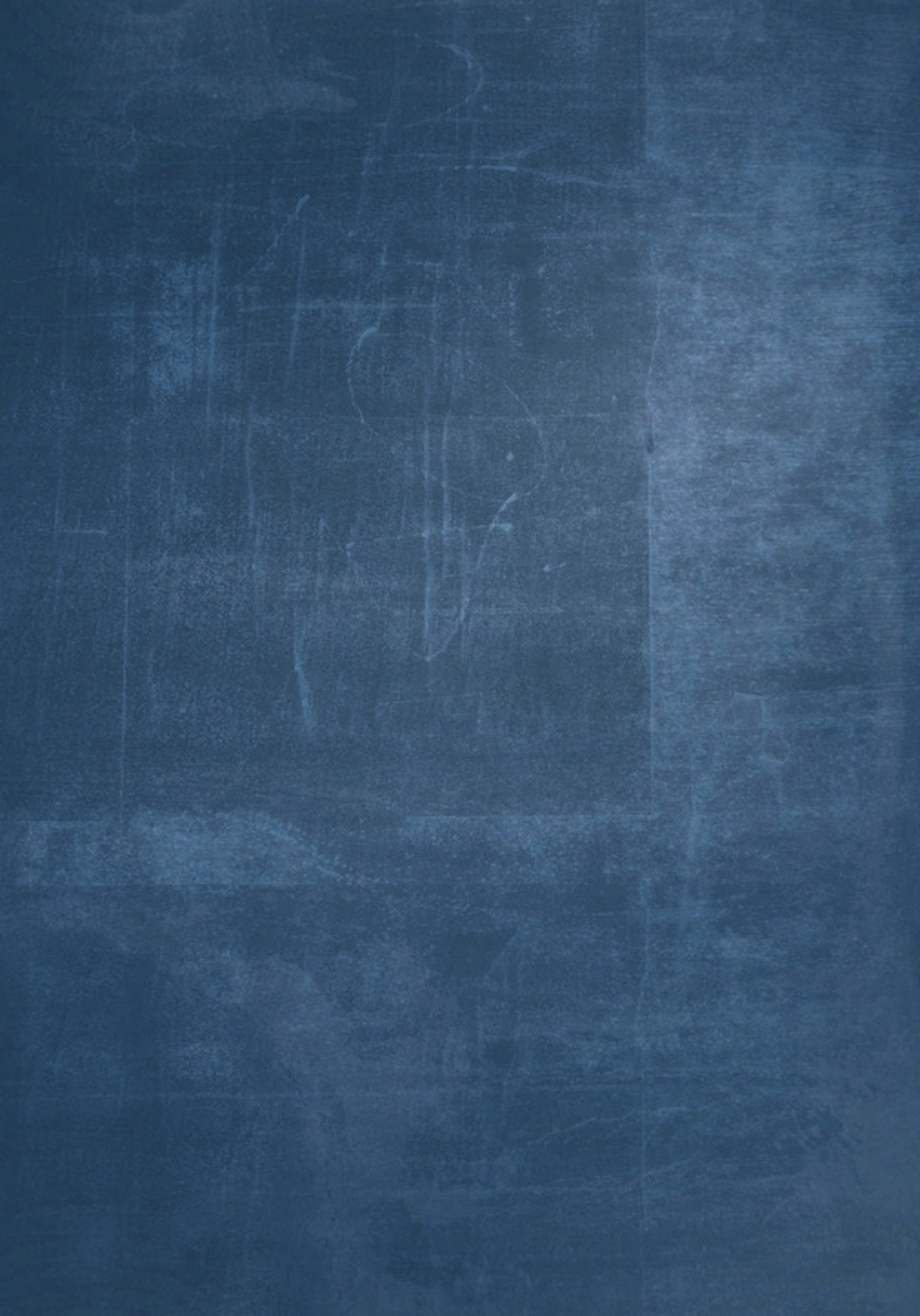}}}

\usepackage[natbib=true,hyperref=true,style=authoryear]{biblatex}

\usepackage{hyperref}
\usepackage{hyperxmp}

\definecolor[named]{OurGreen}{rgb}{0,.392,0}
\AtEndPreamble{
  \hypersetup{
    colorlinks,
    linkcolor=OurGreen,
    citecolor=OurGreen,
  }
}

\title[Catalogs of C and Python Antipatterns by CS1 Students]
      {Catalogs of C and Python\\Antipatterns by CS1 Students}

\author[Bosse, Y. \textit{et al.}]%
       {Yorah Bosse, Igor Scaliante Wiese, Marco Aurélio Graciotto Silva,
        Nelson Lago, Leônidas de Oliveira Brandão, David Redmiles,
        Fabio Kon, Marco A. Gerosa}

\begin{document}


\newgeometry{top=58pt, bottom=44pt, inner=81pt, outer=81pt}
\makeatletter
\BgThispage
\thispagestyle{empty}
\begin{center}
  \null
  \vspace{1\baselineskip}

  \color{white}
  \sffamily

  {\Huge\textsc{\@title}\par}

  \bgroup
  \Large
  \vspace{1\baselineskip}

  Technical Report RT-MAC-2021-01\\
  Department of Computer Science -- IME\\
  University of S\~ao Paulo\par
  \egroup

  {\Huge\vspace{3\baselineskip}}

  \bgroup
  \raggedright
  \large
  \textbf{\textsc{Yorah Bosse}} (yorah.bosse@ufms.br)\\
  {\normalsize\itshape\hspace{1em}Federal University of Mato Grosso do Sul (Brazil)}\\[.5\baselineskip]
  \textbf{\textsc{Igor Scaliante Wiese}} (igor@utfpr.edu.br)\\
  {\normalsize\itshape\hspace{1em}Federal Univ.of Technology - Paraná (Brazil)}\\[.5\baselineskip]
  \textbf{\textsc{Marco Aurélio Graciotto Silva}} (magsilva@utfpr.edu.br)\\
  {\normalsize\itshape\hspace{1em}Federal Univ.of Technology - Paraná (Brazil)}\\[.5\baselineskip]
  \textbf{\textsc{Nelson Lago}} (lago@ime.usp.br)\\
  {\normalsize\itshape\hspace{1em}University of São Paulo (Brazil)}\\[.5\baselineskip]
  \textbf{\textsc{Leônidas de Oliveira Brandão}} (leo@ime.usp.br)\\
  {\normalsize\itshape\hspace{1em}University of São Paulo (Brazil)}\\[.5\baselineskip]
  \textbf{\textsc{David Redmiles}} (redmiles@ics.uci.edu)\\
  {\normalsize\itshape\hspace{1em}University of California (USA)}\\[.5\baselineskip]
  \textbf{\textsc{Fabio Kon}} (kon@ime.usp.br)\\
  {\normalsize\itshape\hspace{1em}University of São Paulo (Brazil)}\\[.5\baselineskip]
  \textbf{\textsc{Marco A. Gerosa}} (marco.gerosa@nau.edu)\\
  {\normalsize\itshape\hspace{1em}Northern Arizona University (USA) / University of São Paulo (Brazil)}\\[1em]
  \egroup

  \bgroup
  \large
  \vspace{7\baselineskip}
  \textsc{2021}
  \egroup

\end{center}
\makeatother

\restoregeometry
\clearpage

\renewcommand\contentsname{}
\thispagestyle{empty}

\makeatletter
\addtocontents{toc}{\protect\setcounter{tocdepth}{-1}}

\begin{center}\sffamily\Huge\color{ACMDarkBlue}Contents\end{center}

\bgroup
\hypersetup{hidelinks}
\tableofcontents
\egroup
\addtocontents{toc}{\protect\setcounter{tocdepth}{2}}
\makeatother

\clearpage

\maketitle

\begin{abstract}
Understanding students' programming misconceptions is critical. Doing so depends on identifying the reasons why students make errors when learning a new programming language. Knowing the misconceptions can help students to improve their reflection about their mistakes and also help instructors to design better teaching strategies. In this technical report, we propose catalogs of \textit{antipatterns} for two programming languages: C and Python. To accomplish this, we analyzed the codes of 166 CS1 engineering students when they were coding solutions to programming exercises. In our results, we catalog 41 CS1 antipatterns from 95 cataloged misconceptions in C and Python. These antipatterns were separated into three catalogs: C, Python, and antipatterns found in code using both programming languages. For each antipattern, we present code examples, students' solutions (if they are present), a possible solution to avoid the antipattern, among other information.
\end{abstract}

\vspace{1\baselineskip}


\section{Introduction}

Learning to program is a complex \citep{mhashi2013difficulties} and arduous task \citep{robins2003learning, jenkins2002difficulty, ala_mutka2003problems},
as evidenced by the number of studies considering the student failure rates shown in the literature \citep{watson2014failure,bosse2016visual,bosse2017programming,bennedsen2019failure,luxton2019pass}.

Understanding the mistakes made by CS1 students when learning to program can provide benefits to the students’ growth. Such mistakes are an essential part of the learning process, and research has shown that students can learn from error \citep{streumer2006world,ellstrom2001integrating}.

In this study, that was a part of the thesis entitled ``Patterns of Difficulties Related to Programming Learning'' \citep{bosse2020thesis},
we observed how a group of CS1 students approached and solved the programming exercises proposed throughout their ``introduction to programming''~class.
Our goal was to identify frequent
misconceptions made by them when developing code and thereby define a set of \textit{antipatterns}.

A pattern is the description of a problem that occurs repeatedly in our environment together with the core of a usable solution to that problem \citep{alexander1977pattern}.
An antipattern, in turn, is a common response that looks like an appropriate and effective solution, but resulting in negative consequences \citep{brown1998antipatterns}.
\citet{brown1998antipatterns}
further add that an antipattern is the result of lack of knowledge or the application of a ``good''~pattern in the wrong context \citep{brown1998antipatterns}.
In addition, they state that mistakes that occurred at least three times in different situations becomes a pattern (antipattern).

In this technical report, we are presenting a set of antipattern catalogs for C, Python, and the intersection of both, in addition to directions on how to learn from them.
These antipattern catalogs may help instructors to teach and students to learn how to program, reducing difficulties faced by CS1 students.
They also may play a key role in research that aims to develop programming oriented teaching-learning systems,
giving greater conditions for developing helpful feedback not only on reducing syntax errors but also on decreasing
semantic errors and misleading programming styles.

\section{Methodology}

The code analyzed in this research came from introductory programming classes taught for engineering courses at the University of São Paulo, Brazil, in 2017. In the following subsections, we detail the steps of data collection and analysis of this study.

\subsection{Data Collection}

The data collection happened in a distance learning course with no face-to-face classes, where all 3 exams were performed on-site. To enroll in this course, students had to have taken the compulsory CS1 face-to-face course in the previous year, had to have at least 70\% attendance, and had to have failed it. There were initially 166 students, 34 of them using the C programming language (Electrical and Computational Engineering students) and 132 students using the Python language (other engineering courses). Each student had to use the same language adopted in their previous face-to-face course when they failed. 

The course took 16 weeks. In each week, students had to solve a set of practical exercises, with automatic evaluation, using Moodle with the VPL plugin
\citep{rodriguez2012virtual}.
Every compilation had the code and its result recorded.
These exercises were mandatory and were considered in the final grade, weighting one fifth of the total grade. 
There were 62 exercises, approximately 4 per week.
The weekly exercises were chosen by the instructor to address concepts of a particular set of topics covering the main topics in programming languages
(e.g., input/output, variables, selection and repetition structure, functions, array). 

\subsection{Data Analysis}

For the analysis, we chose exercises covering all of the content taught in the course. The Appendix \ref{ape:exercises} present the statements of the 15 chosen exercises.
Exercises were grouped by topic and we chose representative exercises giving preference for the first ones in each topic. From the 15 chosen exercises, we analyzed 2,233 student code submissions (1,253 in C and 980 in Python). All students, $\sim$20\% using C and $\sim$80\% using Python, solved the same exercises. 

After choosing which exercises to explore, the choice of code submissions to be analyzed for each of them was random, but giving preference to students' codes with up to 20 submissions. For each exercise, a minimum of 50 submissions was analyzed. 

During the analysis, several misconceptions were found. Each new mistake found was cataloged in a new datasheet, and those that already existed were added as a new event in a existing one. 

We used the following criteria to classify the 3 types of errors: syntax, semantics or style.
\textbf{Syntax errors} detected by the compiler; \textbf{semantic errors} identified by the wrong results;
and the \textbf{Style errors} considered not appropriate because they could generate future problems in code development or poor programming practice \citep{rogers2014aces}.

When we finished analysing the codes and classifying the mistakes, the antipattern catalogs of each language were organized with the errors that presented themselves in at least
three distinct student events. Next, we organized a catalog with the antipatterns that appear in both languages.

\section{Antipatterns Catalogs}

Codes developed in C and Python were recorded and analyzed during CS1 taught for engineering courses at the University of São Paulo, Brazil.
We found 95 types of misconceptions made by students using C and 44 by students using Python.

The misconceptions with at least three registered events yielded three antipattern catalogs (Table~\ref{tab:antipatterns_table}).
The first catalog presents the antipatterns found in codes developed with the C programming language, a total of 21 (rows 1 to 21).
The second catalog presents the antipatterns found in Python, totaling 11 (rows 31 to 41). The third one presents the 9 that occurred in both C and Python programming languages (rows 22 to 30). 

\begin{table*}
  \caption{Summary of the antipattern catalogs.}
  \label{tab:antipatterns_table}
  \includegraphics[height=\textheight, width=\textwidth]
  {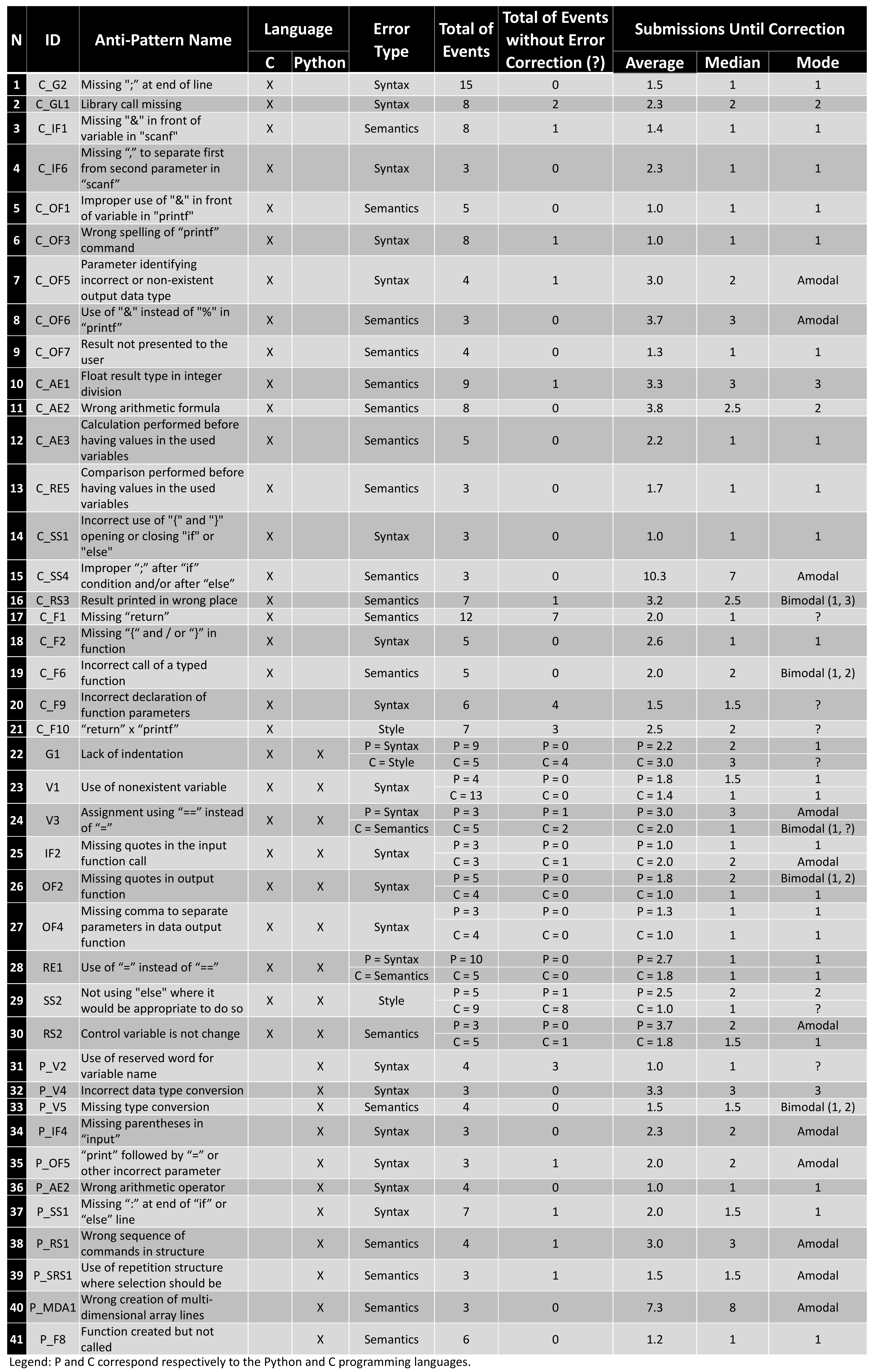}
\end{table*}

The last five columns of Table~\ref{tab:antipatterns_table} show the total of events found, in how many of these events the student was unable to solve the problem and the average, median, and mode of submissions necessary to correct the error. Appendices \ref{ape:antipatterns_C}, \ref{ape:antipatterns_Python}, and \ref{ape:antipatterns_C_Python} show the complete datasheets of all antipatterns that make up each catalog, according to the format explained in the Figures \ref{fig:part1}, \ref{fig:part2}, and \ref{fig:part3}. 

\subsection{Description of the Antipatterns Format}

\begin{figure}
   \centering
   \includegraphics[width=5.5in]{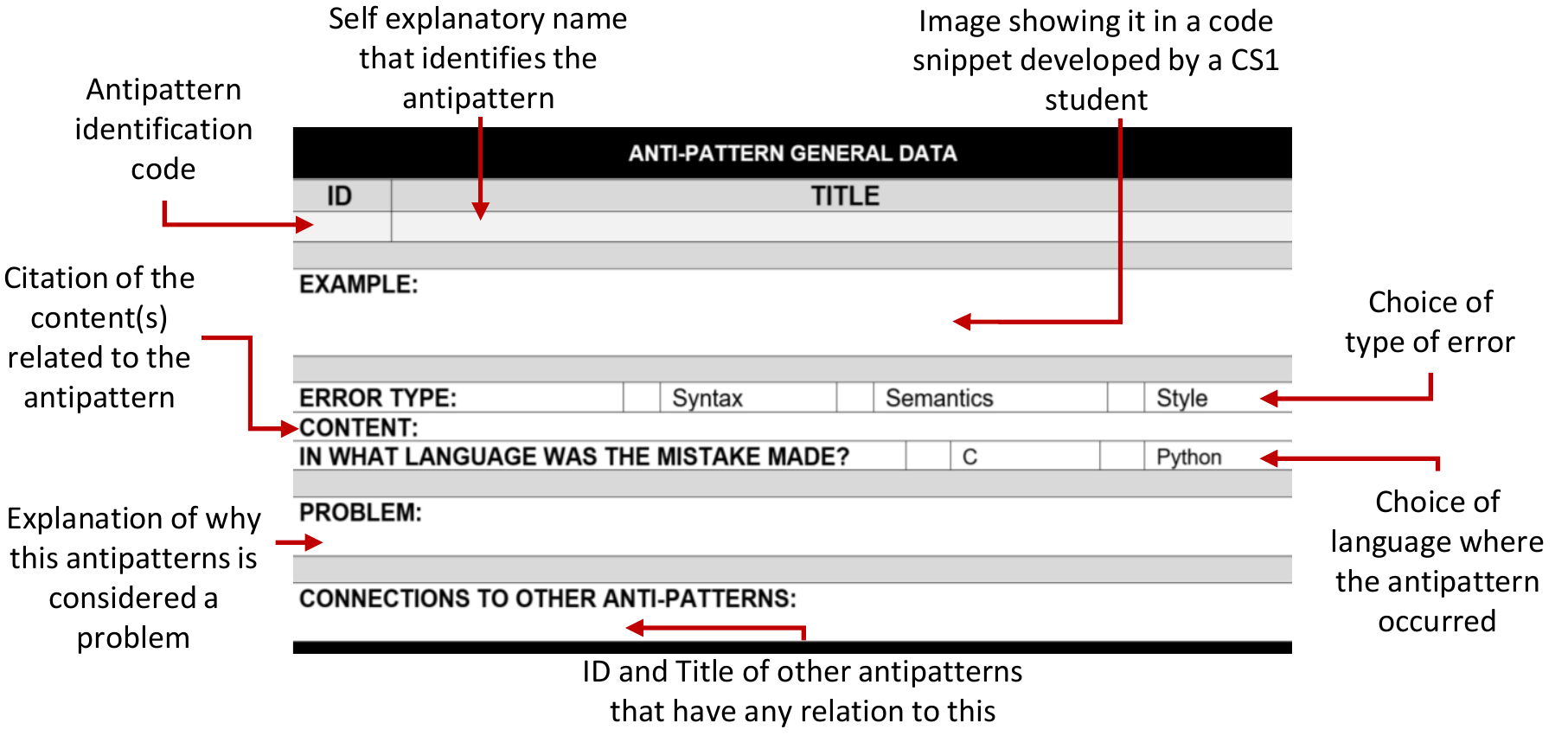}
   \caption{Explanation of the first part of the antipatterns datasheets.}
   \label{fig:part1}
\end{figure}

The format adopted for the datasheets is an adaptation of the presentation template of a pattern suggested by~\citet{alexander1977pattern}.
Each datasheet of the catalog consists of three parts. The first part serves to identify the antipattern in a unique way. Figure~\ref{fig:part1} shows the information that makes up this part of the datasheets. 

\begin{figure}
   \centering
   \includegraphics[width=5.5in]{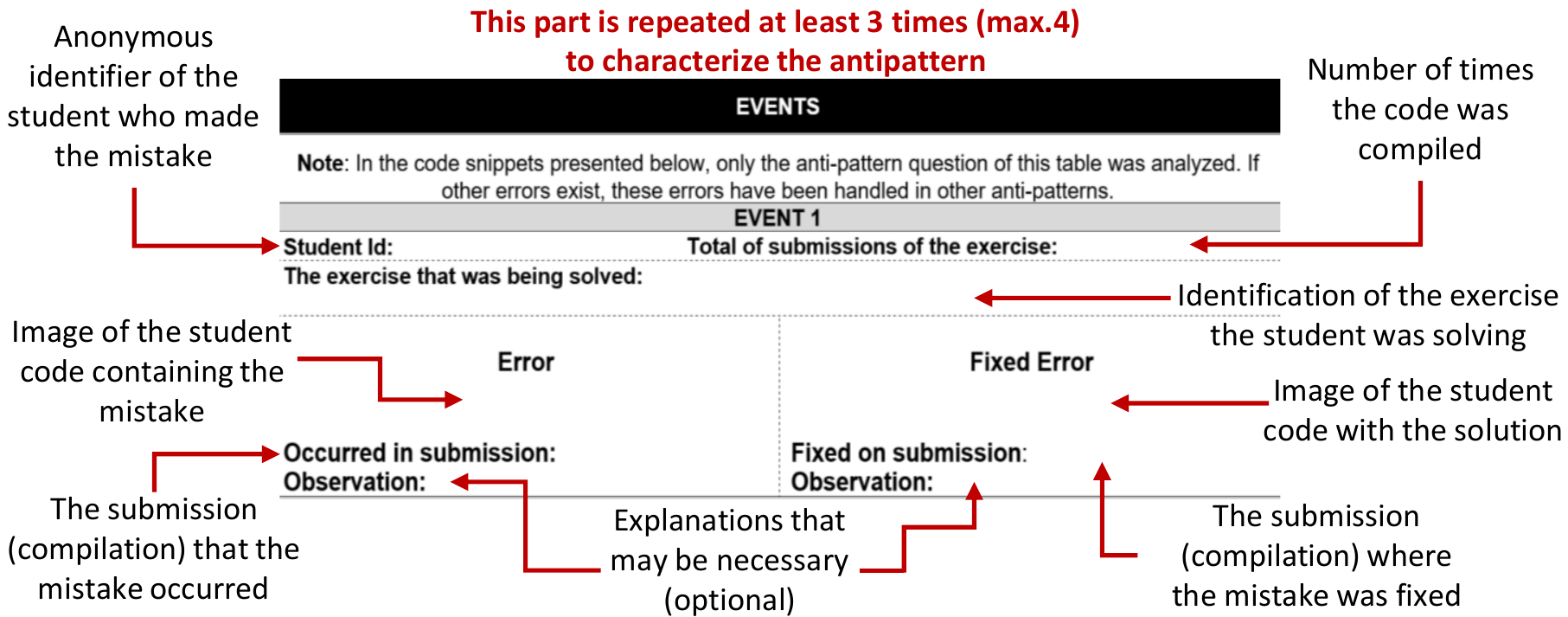}
   \caption{Explanation of the second part of the antipatterns datasheets.}
   \label{fig:part2}
\end{figure}

The second part is intended to show in detail the occurrences found. There are three to four events presented to characterize an antipattern.
Information such as in which submission/compilation the mistake occurred and in which it was solved helps, for example, to know if the student solved the antipattern and, if so, to show how and how much 'time' was necessary for that (Figure~\ref{fig:part2}).
Appendix~\ref{ape:exercises} shows the statements of the exercises identified by item ``The exercise that was being solved.'' 

\begin{figure}
   \centering
   \includegraphics[width=5.5in]{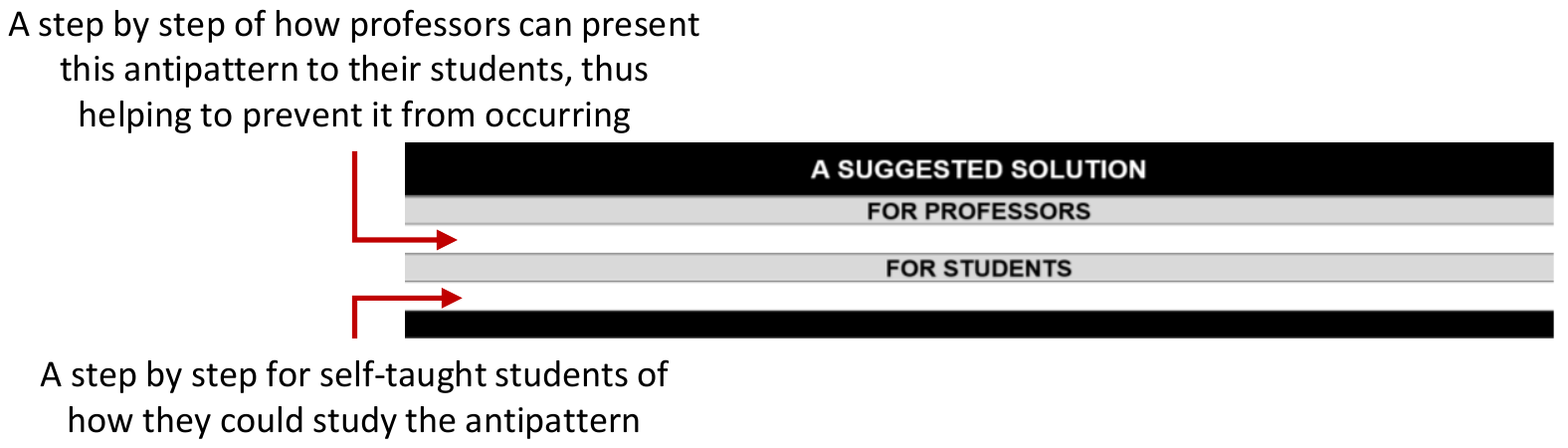}
   \caption{Explanation of the third part of the antipatterns datasheets.}
   \label{fig:part3}
\end{figure}

The last part serves to suggest a step by step of how professors can present the antipattern to their students, and also suggestion of a step by step of how students can study it by themselves (Figure \ref{fig:part3}).
Appendix~\ref{ape:solution_prof_students} shows the statements of the exercises identified by item
``The exercise that was being solved.'' 

\section{Final Consideration}

We found 139 types of misconceptions in C and Python, which 41 of them were identified as antipatterns in the CS1 students' code. Professors could work these antipatterns while teaching the corresponding content, avoiding the time spent looking for and solving problems in the codes developed by their students.

The data exposed in the last five columns of Table~\ref{tab:antipatterns_table} can provide vital information for professors. With this data, the professor can identify the antipatterns that most occurred, those that students typically failed to resolve, those that took a long time for students to find a solution, among much other information.

These catalogs can be another step in a broader movement to lift and categorize the specific difficulties faced by students who may help professors to teach and students to learn programming, as well as researchers in the development of new research and tools to assist them in these tasks.

\subsection{Future Work}
As a future work, a study that verifies the \textbf{impact on the CS1 students' code in case the antipatterns were constantly highlighted} during the introductory programming classes. Besides, the \textbf{creation of a questions repository} could be helpful. This repository's goal would be to address each antipattern in many different ways, looking to develop some essential skills for programmers, like finding errors in their code.
According
to~\citet{gomes2014teacher}, one way to help students is with well-designed, carefully thought out exercises, reflecting how, which, when, and why each exercise would be presented. A question repository could help the professors when choosing and varying the exercises used in each situation.

\section*{Acknowledgment}
\phantomsection
\addcontentsline{toc}{section}{Acknowledgement}

The first author was supported by a scholarship from the CAPES Foundation, an agency under the Ministry of Education of Brazil -- Program name: PDSE--Programa de Doutorado Sanduiche no Exterior (Doctoral Program Abroad)/Process number: PDSE -- 88881.135066/2016-01), to study in the University of California -- Irvine, and also supported by UFMS -- University of Mato Grosso do Sul to develop the thesis. In addition, this study was financed in part by the Coordena\c{c}\~{a}o de Aperfei\c{c}oamento de Pessoal de N\'{i}vel Superior -- Brazil (CAPES) -- Finance Code 001.

\printbibliography[heading=bibintoc]

\appendix
\section{Statements of the Chosen Exercises}
\label{ape:exercises}

Below, we present the statements of the exercises chosen to analyze the codes generated by the students of introduction to programming.

\begin{exercises}
\item[Exercise 1.1 --] Develop a code that reads an integer entered by the user and prints the value typed.

\item[Exercise 1.2 --] Develop a code that receives an integer (int), typed by the user, and prints the square of that number.

\item[Exercise 2.1 --] Develop a code that reads two integers typed by the user, and prints the largest of the two numbers.

\item[Exercise 2.2 --] Develop a code that reads two integers as inputs (say in variables $a$ and $b$), and prints 1 if ``$a < b$'' and $0$ otherwise.

\item[Exercise 3.1 --] Develop a code in C or Python that checks whether 3 numbers can represent the angles of a triangle or not. Your algorithm should prompt the user to enter the 3 natural numbers (representing angles in degrees) and print: ``Sim'' (Yes) and the 3 numbers in the sequence entered if they add up to 180, and ``NAO'' (NO) and the sum of the 3 numbers otherwise.

\item[Exercise 3.2 --] Develop a code that reads three natural numbers as inputs and checks to see if those numbers are Pythagorean. Three numbers are Pythagorean if the square of the
largest of them (hypotenuse) is equal to the sum of the square of the other two. Your program should print: if Pythagorean, the value ``1'' and the value of the hypotenuse squared; if not Pythagorean, only the value ``0''.\par
Three numbers, represented by $h$, $a$, and $b$ are called Pythagorean if, and only if, $h^2 = a^2 + b^2$.
The name is because they correspond to the sides length of a right triangle.

\item[Exercise 3.3 --] Develop a C/Python program that reads a natural (say in the "total"
variable) and computes the largest sum of consecutive naturals, starting at 1, that is less
than or equal to the value entered ($<=$ total).

\item[Exercise 4.1 --] Develop a program with functions named ``soma'' (sum) and ``media''
(average), both with 2 formal parameters of type ``int'', which respectively return the sum
of the parameters and their average. Functions must have 2 integer parameters, such as
``soma(a, b)'' and ``media(a, b)''. Your main program should receive from the user 2 integers,
invoke with them first the sum function and print their sum, then invoke media and print
their arithmetic mean (invoking the functions).

\item[Exercise 4.2 --] Make a program in C~/~Python necessarily implementing a function named
$fat$, with one parameter $n$, which, given an integer $n$, returns the factorial of $n$ if $n \geq 0$, otherwise return $-1$.
Make a "main" program that prompts the user to type any integer and using the function $fat$ returns the factorial of $n$, if $n\geq 0$, otherwise return $-1$.\par
Remember, the factorial function $fat(n)$ is defined only for naturals (starting by zero), as follows:
$fat(0) = 1$ and $fat(n) = n * fat(n-1), n>0$. 
Thus, for example, the factorial of $4$ (represented by $4!$) is computed as follows: $fat(4) = 4! = 4.3.2.1 = 24$.\par
Therefor, it is necessary that your program prints $-1$ whenever the user enter a negative integer (e.g., if the input is $n = -10$, the output is $-1$; input $n = -21$ implys output $-1$).

\item[Exercise 4.7 --]
Develop a code that receives a positive natural number $n$ ($n \geq 1$) and then
$n$ lists of integer values, each list is ended by a $0$ (ending mark of the subsequence),
calculating the sum of all odd numbers in each subsequence and print them (as soon as you read each $0$).\par
Examples: For the subsequences below, all with $n = 3$\par
\begin{verbatim}
Ex. 1. Inputs (including n): 3 0 0 0 => outputs: 0
Ex. 2. Inputs (including n): 3 2 0 2 0 2 0 => outputs: 0
Ex. 3. Inputs (including n): 3 1 0 3 0 5 0 => outputs: 1 3 5) \end{verbatim} 

\item[Exercise 7.1 --] Construct an algorithm that receives a positive natural n ($n > 0$) and prints 
the sum of the first n terms of the harmonic series defined below.\par
$H = 1/1 + 1/2 + 1/3 + 1/4 + \dots\space + 1/k + \dots$ 

\item[Exercise 7.2 --] Build an algorithm that takes a positive natural number and real value ($n$ and $x$),
calculates and prints the value of $x\hat\ n$ (power from $x$ to $n$). Use only the sum and product operators.

\item[Exercise 7.4 --] Build a program that receives an integer $n$ and then $n$ characters as input
by printing the ASCII (American Standard Code for Information Interchange) code for each character.

\item[Exercise 8.1 --] 1. Implement a function called \verb|determineSeOrdenado| that takes as parameters an integer $n$ and array of integers called $vet$.
The function should return $1$, if the elements in \verb|vet[]| are in ascending order and $0$, otherwise
(that is, if \verb|vet[0] < vet[1] <| \dots\space \verb|< vet[n-1]|, returns $1$ and in any other setting returns $0$).\par 
2. Make a \verb|main| program in which the user types a number \verb|n>0|, followed by $n$ integer 
values that are stored in the array (which can have up to $40$ elements).
Using the value returned by the function of item 1, your program should print $1$, if the array is in ascending order, and $0$ otherwise.

\item[Exercise 13.1 --] Make a program in which the user enters positive integers $m$ and $n$
(\verb|m > 0, n > 0|), followed by $m$ x $n$ real values (float), which should be stored in a matrix of order $m$ x $n$.
Then your program should print each line, followed by the ``\verb|=|'' sign and the sum of the line, starting at line 0, then line 1, and so on.
Thus, you must first print the value of \verb|M[0][0]|, followed by \verb|M[0][1]|, and so on to the last of the line (\verb|M[0][n-1]|) and then 
the equal sign and the sum of the number from line zero. Then follow lines 1, 2, 3 to m-1.

\end{exercises}

\pagebreak


\section{Suggested Solution for Professors and Students}
\label{ape:solution_prof_students}

Below, we present detailed descriptions of solutions for professors and students mentioned in the antipattern catalogs.

\setlength\parindent{0pt}
\setlength\parskip{1\baselineskip}

\textbf{FOR PROFESSORS}

\textbf{1. Explanation using blackboard and projector -- reinforce the concept using
Kahoot}

It is important to make clear to the learner that this anti-pattern exists. For this, a solution
suggestion would be:

\begin{enumerate}
\item Bring to class code snippets with antipatterns related to previously presented topics
with at most one error per line
\item show them to the learners
\item ask them to point out some error
\item ask if someone in class could explain the consequences of the error for the program
\item reinforce the explanation, if possible, showing other examples of the same error
\item explain the correct way to write the command (line), without the error
\item rewrite the line on the blackboard, without the error, or make the correction visible
on the projector
\item repeat steps 3 to 7 as long as there are errors in the code
Then (it is necessary for the teacher’s computer and the learners’ phones or
computers to have Internet access):
\item show the password for the previously prepared kahoot quiz and ask learners to con-
nect to kahoot with their phone
\item If this method has not been used before with them, explain the “game”
\item start the “game”, cheer each learner’s victory, and make notes on the questions they
fail most often
\item In the end, present the results and go back to the exercises they had the most diffi-
culty with.
\end{enumerate}

P.S: The teacher will obtain a valuable resource, as it will become clear on which topics
students need reinforcement and may address them further during class.

\pagebreak

\textbf{2. Groups working with step-by-step execution to understand errors --
reinforce with exercises}

It is important to make clear to the learner that this anti-pattern exists. For this, a solution
suggestion would be:

\begin{enumerate}
\item assemble teams of 2 or 3 students;
\item give the teams codes with the error. Not necessarily the same code for all teams;
\item ask them to do a step-by-step execution of the code presented;
\item then ask for a team to show the class what they have found and explain how they
got this result;
\item ask others with the same snippet if they had different results. If so, ask them to
present to the class;
\item now you (instructor), using visuals, simulating computer memory space, for in-
stance, show, correct or reinforce what was presented;
\item repeat steps 3 through 6 if you have more than one type of exercise;
\item give them exercises to solve in which they could make the mistakes mentioned
above;
\item ask them to solve on the blackboard so you can correct them if the mistake still
appears.
\end{enumerate}

\textbf{3. Programming in front of the students, making the error appear-- reinforce
by asking students to develop new codes}

It is important to make clear to the learner that this anti-pattern exists. For this, a solution
suggestion would be:

\begin{enumerate}
\item ask learners to solve an exercise in which the error you want to address might
happen;
\item go to the blackboard and expose a superficial outline of what the code needs to do
and in what sequence;
\item use the projector to start programming on the computer in front of the students;
\item make the error appear;
\item compile and ask the learners for suggestions on how the code could be fixed;
\item test each suggestion and explain if necessary, the reasons why it did not work;
\item finish with working code projected for learners to see;
\item give them one more exercise to solve;
\item ask a learner to develop their code right on the computer to project for all;
\item explain what is happening while the learner is typing;
\item repeat steps 8 through 10 as many times as you like, using intensive practice for
learning.
\end{enumerate}

\pagebreak

\textbf{4. Apply the bench test (table test) in code samples with and without the
error, compare the results -- reinforce by asking the students to solve
some exercises}

It is important to make clear to the learner that this anti-pattern exists. For this, a solution
suggestion would be:

\begin{enumerate}
\item show an example of the error to the learner. Make sure the only mistake in your
example is the one you want to address;
\item make it very clear, highlighting the mistake with a colored circle or arrow;
\item explain why it is wrong;
\item do a bench test (table test) showing what happens;
\item fix the code, and redo the bench test (table test);
\item compare the results;
\item make sure the learners understand by asking them to do one or two exercises like
this, but with other code snippets.
\item correct during class, in the presence of the learners.
\end{enumerate}

\textbf{FOR STUDENTS}

\textbf{1. Solve exercises, add code errors, and ask classmate to find them}

Learners may study this antipattern to avoid it in the future with an exercise:

\begin{quote}
\underline{\strut Note:}\itshape It is important that students work with at least one peer and that they have a
computer to work on the exercise.
\end{quote}

\begin{enumerate}
\item Study some of the antipatterns of a given topic
\item Choose similar exercises that depend on the topic and solve them independently
\item After the exercises are done and error-free with correct outputs, each one copies
their code to paper with added errors (each one chooses how many errors to add)
\item Switch papers
\item Each one corrects their peer’s exercise, pointing out the errors found
\item When done, learners check with their peer, who previously solved the exercise,
whether any problem slipped through or if all errors were found
\item Each one explains the solution they developed to their peer, showing the difficulties
they had and how they solved them.
\end{enumerate}

\pagebreak

\textbf{2. Study one or more anti-patterns, introduce to classmate and reinforce
with exercise}

Learners may study this antipattern to avoid it in the future with an exercise:

\begin{quote}
\underline{\strut Note:}\itshape It is important to do this with at least one peer.
\end{quote}

\begin{enumerate}
\item Choose a topic already seen in class and select different parts of the topic for each
learner to explain
\item Each one is to study the material, prepare a quick presentation, and search for
examples to illustrate the explanation
\item Search for at least 1 exercise to be solved after your presentation
\item Choose whose turn it is to present
\item Start the presentation. The explanation should clarify everyone’s difficulties.
Showcase the possible errors (antipatterns) that may occur. In case some doubt
remains, all may help search for answers if the presenter does not know enough.
\item Present the chosen exercise and ask colleagues to solve it
\item Check it together with bench test (table test) of the found solutions
\item Analyze what would happen if the antipattern existed in the code
\item Repeat steps 4 to 9 as many times as necessary for all to present their part
\end{enumerate}

\textbf{3. Introduce the error into a code and understand the consequences it
generates.}

Learners may study this antipattern to avoid it in the future with an exercise:

\begin{quote}
\underline{\strut Note:}\itshape It is important to have a computer available to undertake the exercise.
\end{quote}

\begin{enumerate}
\item Copy into the development environment some completed exercise from a previous
class, a book, or the internet where the error (antipattern) may be inserted
\item execute the program and reason about what it does
\item insert the error (antipattern) in the code and try to execute it again
\item If there is a compilation error, note how the compiler present the error, so you can
more easily understand what the compiler points to in cases of error
\item If the program executes, note how results are presented, trying to understand where
the error is and how it behaves
\item Do this with multiple exercises, until you are certain that you understand the
antipattern
\end{enumerate}

\clearpage

\section{Antipattern Catalog in C}
\label{ape:antipatterns_C}

Below, we present the catalog of antipatterns found in code developed using C as programming language.

\begin{minipage}{\textwidth}
\includepdf[scale=1,pages=1,pagecommand={}]{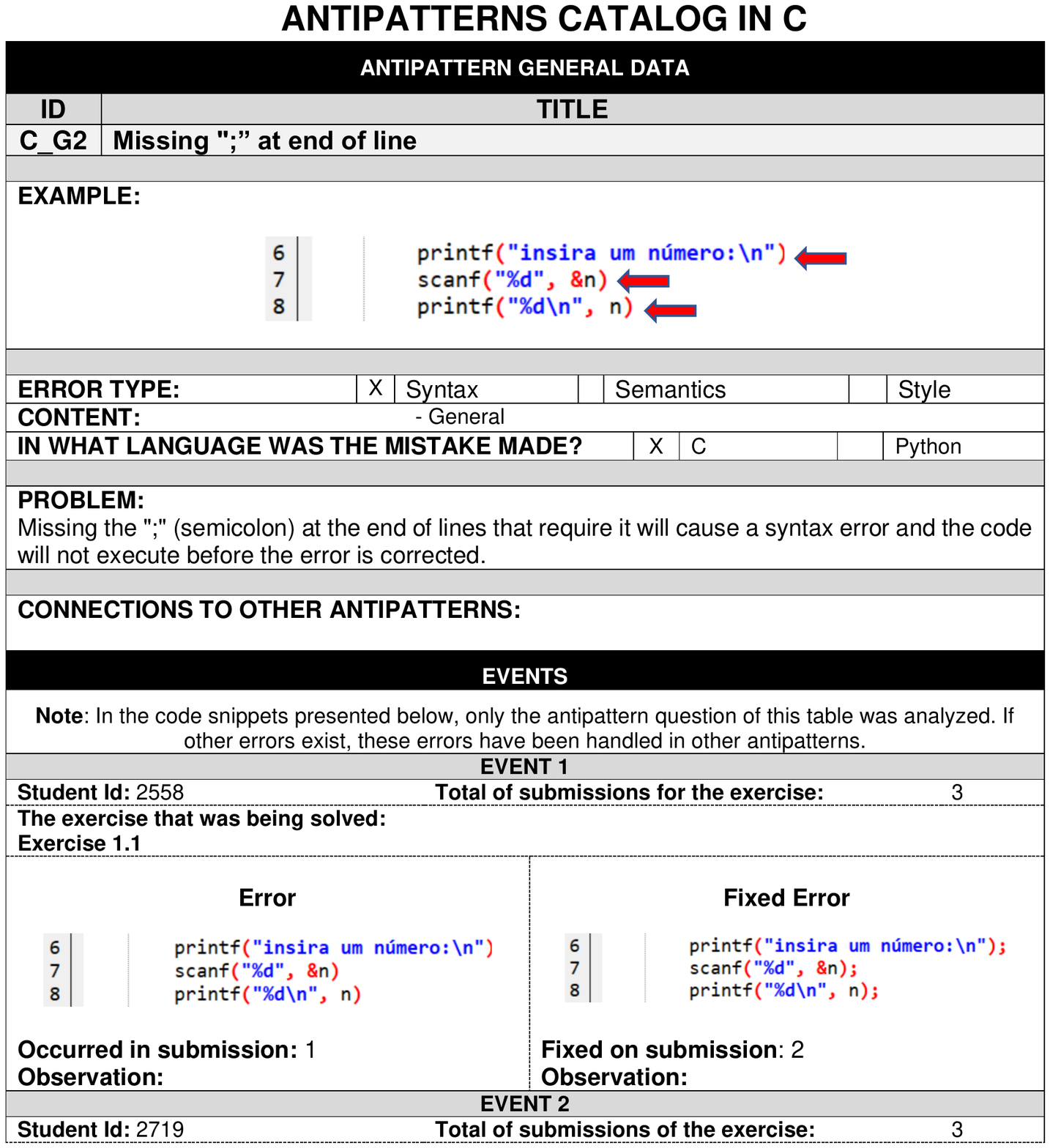}
\end{minipage}

\includepdf[pages=2-,pagecommand={}]{./appendix/antipatterns_catalog_c.pdf}

\pagebreak

\section{Antipattern Catalog in Python}
\label{ape:antipatterns_Python}

Below, we present the catalog of antipatterns found in code developed using Python as programming language.

\begin{minipage}{\textwidth}
\includepdf[scale=1,pages=1,pagecommand={}]{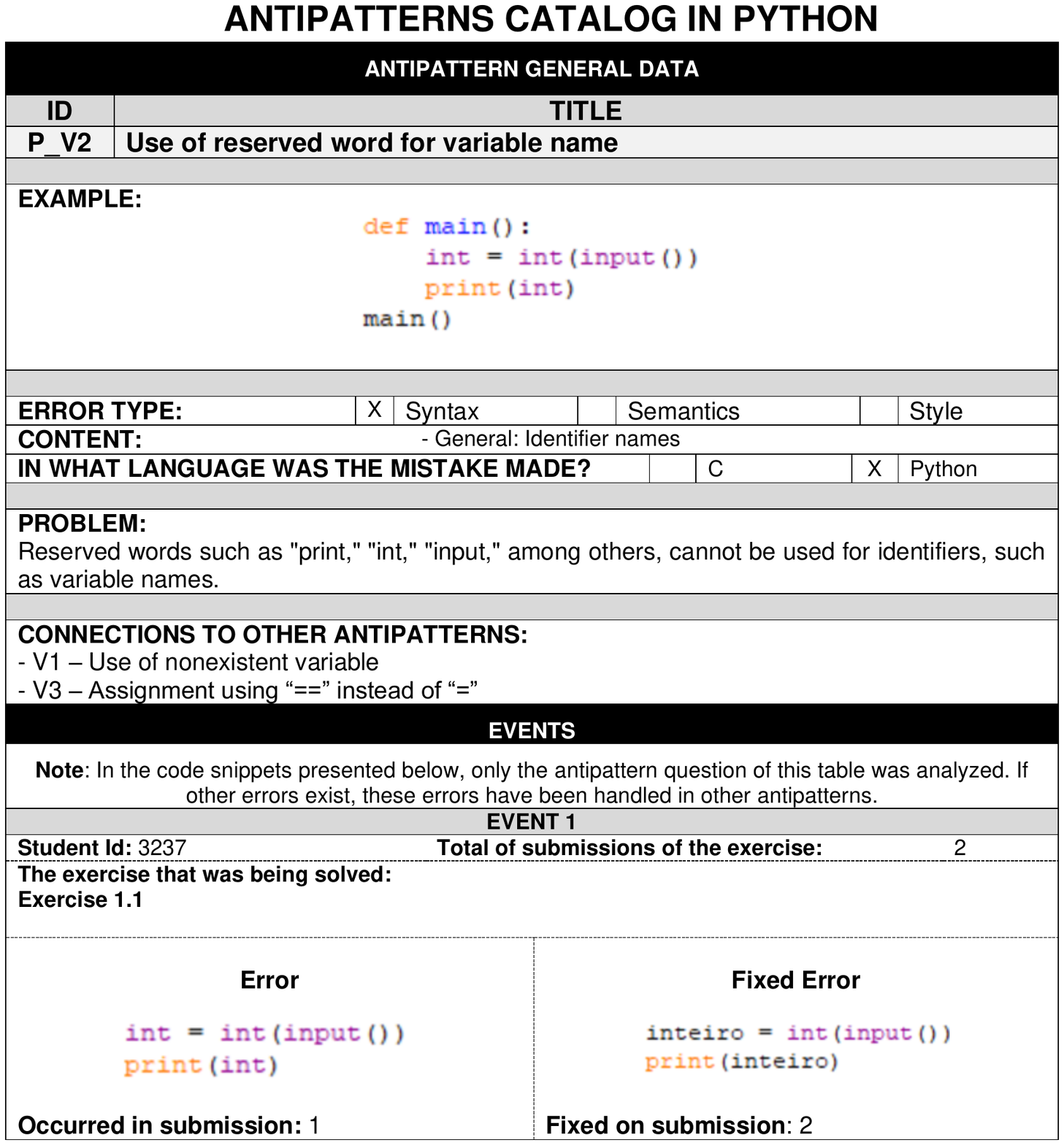}
\end{minipage}

\includepdf[pages=2-,pagecommand={}]{./appendix/antipatterns_catalog_python.pdf}

\pagebreak

\section{Antipattern Catalog in Both, C and Python}
\label{ape:antipatterns_C_Python}

Below, we present the catalog of antipatterns found in code developed using Python and C as well. 

\begin{minipage}{\textwidth}
\includepdf[scale=1,pages=1,pagecommand={}]{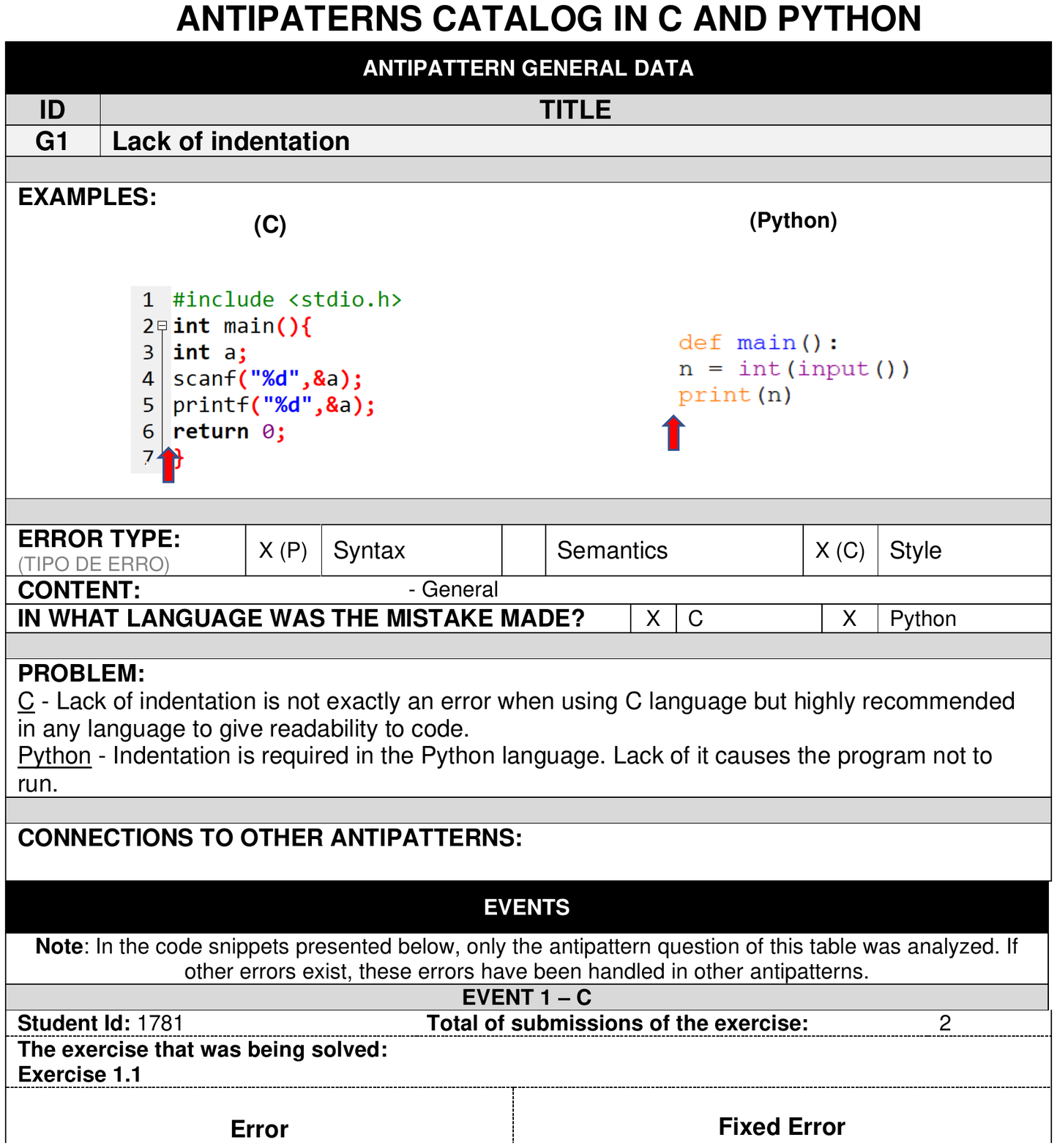}
\end{minipage}

\includepdf[pages=2-,pagecommand={}]{./appendix/antipatterns_catalog_c_python.pdf}

\end{document}